\documentclass{iopconfser}
\usepackage{graphicx}
\usepackage{aas_macros}
\usepackage{lineno}

\begin{document}

\title{FRAM Next Generation at the Pierre Auger Observatory: cloud monitoring in the age of CMOS cameras}

\author{J Ebr$^{1}$ for the Pierre Auger Collaboration$^{2}$ and S Karpov$^{1}$}

\affil{$^1$FZU -- Institute of Physics of the Czech Academy of Sciences, Prague, Czech Republic}
\affil{$^2$Observatorio Pierre Auger, Av. San Mart{\'i}n Norte 304, 5613 Malarg\"{u}e, Argentina}
\affil{Full author list:  http://www.auger.org/archive/authors2024\_07.html}    

\email{ebr@fzu.cz}

\begin{abstract}
The visibility of stars is often used for cloud detection using all-sky cameras, which have however only a limited reach and resolution near the horizon due to the lack of detectable stars. At the Pierre Auger Observatory, it is also used by the current generation of FRAM robotic telescopes, but -- due to their limited field of view -- only for a small number of selected showers. Thanks to the recent development in astronomical CMOS cameras, we are able to propose a new type of device, specifically tailored to the field of view of the fluorescence detectors (FD) of the Pierre Auger Observatory. The sub-second readout times available with CMOS cameras allows the efficient use of short exposures, and so the field of view of one FD can be covered within half a minute with a resolution and reach sufficient to detect small clouds with a setup that is significantly smaller, simpler and cheaper than the current FRAMs. The FRAM Next Generation (framNG) device will be able not only to detect clouds, but also to assess their optical thickness, provide information on aerosol extinction, sky brightness and possibly even record atmospheric phenomena and astrophysical transients. The main challenge lies in the large data volume produced which necessitates reliable real-time data processing.
\end{abstract}

\section{Motivation}
The Pierre Auger Observatory \cite{PierreAuger:2015eyc} employs several complementary methods to ensure that the longitudinal profiles of air showers recorded by the Fluorescence detector (FD) are not significantly affected by the presence of clouds. These methods have so far included Lidars and infra-red (IR) cameras installed at each of the four FD stations, two laser facilities in the center of the array and the use of data from geostationary satellites \cite{Chirinos}. While all of these methods contribute to a complex view of the clouds above the Observatory, only the IR cameras have provided a detailed image of the momentary distribution of the clouds in the FD field of view; however their operation has been recently discontinued for technical reasons and a new method of cloud detection, based on the sky brightness measured directly by the FD, is being developed \cite{bianca}.

The knowledge of the detailed distribution of clouds in the field of view is particularly important for the study of showers with anomalous longitudinal profiles \cite{Blazek} as the presence of clouds can distort the longitudinal profile, thus mimicking a truly anomalous profile. The FRAM robotic telescopes \cite{2021JInst..16P6027A} conduct follow-up observations of showers identified as candidates for having such anomalous longitudinal profiles. They take a series of images along the apparent track of each shower and detect clouds using stellar photometry. However, due to their limited field of view, they can provide cloud information only for a small subset of showers given the available observation time. Moreover, the FRAMs, being full telescopes with equatorial mounts, are relatively expensive to build and maintain, and thus they have been so far installed only at two of the four FD stations.

Recent advances in astronomical cameras, in particular in the use of CMOS detectors in astronomical applications \cite{2020SPIE11454E..0GK}, allow us to propose a new FRAM-like device able to provide detailed cloud information based on stellar photometry for every shower, at a cost which makes it feasible to install at all four FD stations. The availability of other cloud monitoring devices and methods on the Auger site will enable us to quantify the performance of the device and to optimize data processing algorithms.

\begin{figure}
\begin{center}
\includegraphics[width=37pc]{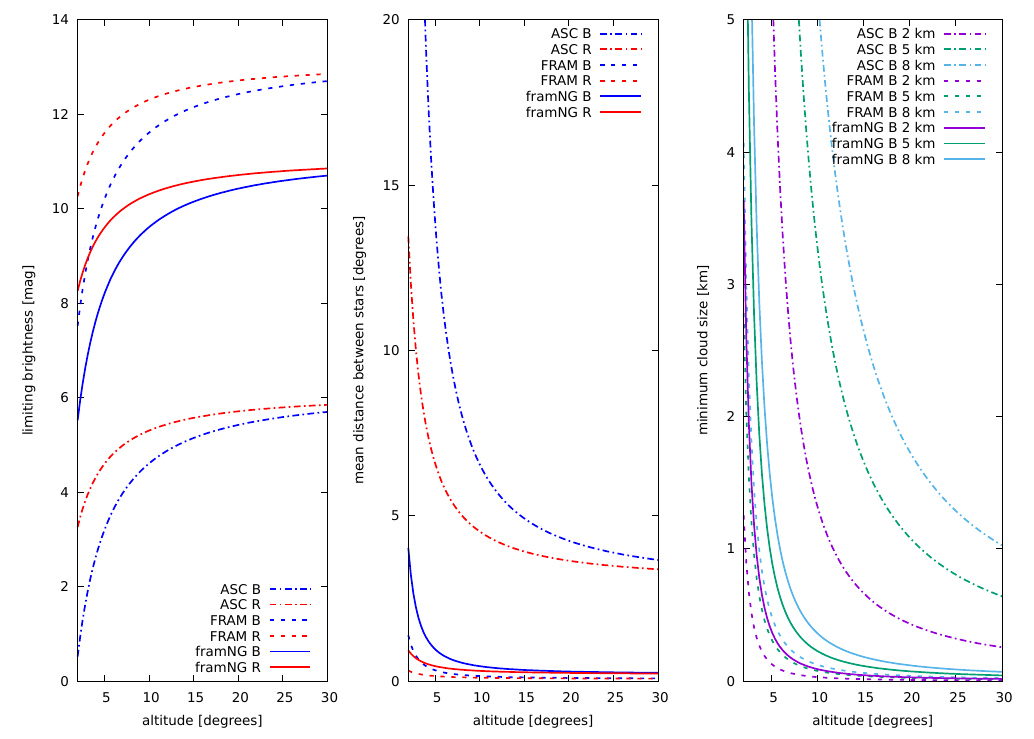}
\end{center}
\caption{\label{plot}Comparison of idealized models of three methods of cloud detection -- all-sky camera (ASC), current FRAM telescopes and the proposed framNG system. Left: limiting magnitudes for B and R filter observations during clear nights. Middle: mean distance between detected stars in an image, as calculated from the limiting magnitudes and the average star density on the sky. Right: minimum physical sizes of clouds for three different cloud heights, calculated from the mean distances between detected stars.}
\end{figure}

\section{Cloud detection using stars for the Pierre Auger Observatory}

The field of view of each station of Fluorescence detector of the Pierre Auger Observatory \cite{ABRAHAM2010227} spans 180 degrees in azimuth -- to cover such an area in a single exposure, an all-sky camera is required. Cloud detection for air-shower observatories with all-sky cameras using stars is a well-established technique, in particular for imaging atmospheric Cherenkov telescopes \cite{Mandat,Adam}, but it is less suitable for the purposes of fluorescence detectors, which tend to operate in lower altitudes above the horizon, where less stars can be detected due to atmospheric extinctions. The Auger FD field of view covers altitudes between 2 and 30 degrees above the horizon (with the exception of the Coihueco station, where the HEAT extension reaches up to 60 degrees). At the highest energies, showers at distances over 40 km from the FD can be efficiently detected \cite{PierreAuger:2010swb}. A shower occurring 40 km away from the FD is roughly a kilometer above the surface when it is observed 2 degrees above the horizon. This is a height at which the two crucial conditions can occur simultaneously: many showers can still produce significant fluorescence light and clouds may be present. From this we see that any method to detect clouds must be sensitive down to the lower edge of the FD field of view.

Figure~\ref{plot} presents a simplified comparison between three different idealized tools for star-based cloud detection which differ only in the limiting magnitude in zenith and their performance changes with altitude only due to atmospheric extinction. The presented instruments are:  all-sky camera (6 mag), FRAM (13 mag) and the proposed "FRAM Next Generation" (framNG) device (11 mag); we will explain the technical details of covering the entire FD field of view with such sensitivity in the next section. In the left panel we show for each device the limiting magnitude as a function of altitude above horizon for a typical clear night in two Johnson filters, B and R. For measurements of atmospheric transparency at Auger, B filter is usually preferred, because it is the closest easily available bandpass to the near-UV range in which the FD operates. If the main targets are clouds, which are mainly ``gray'' (with optical depth largely independent of wavelength), the use of R filter can be more practical thanks to lower molecular extinction. Alternatively a filter can be excluded entirely, because the new data from the GAIA satellite \cite{2023A&A...674A...1G} provide a photometric reference in an arbitrary bandpass. 

The middle panel of Figure~\ref{plot} converts the limiting magnitude into the average distance between stars, using the average stellar density across the sky; while in the densest parts of the Milky Way, the distance is even orders of magnitude  smaller, in the sparsest regions of the sky it is typically less than 2 times larger \cite{2013AstBu..68..481Z}. This conversion already shows why traditional all-sky cameras are insufficient for the detection of clouds at lower altitudes, but  it is even more clearly shown on the right panel of Figure~\ref{plot}, where the angular distance is further converted into the physical size of the cloud for three different cloud heights. Not only are there less stars detected at lower altitudes, but the clouds there are much further away and their angular sizes are smaller for the same physical size. Note that at 40 km, the altitudes above horizon are roughly 3, 7 and 11 degrees for clouds at heights of 2, 5 and 8 kilometers. We see that framNG should be able to detect a kilometer-sized cloud in the entire relevant range and much smaller clouds over most of the field of view -- and, more importantly, that the loss of resolution with respect to the current FRAM telescopes is relatively small.

\section{CMOS sensors and the framNG concept}
CCD sensors, which started to be used in astronomy in 1970s, have been the standard imaging tool for many decades now, thanks to their low noise, homogeneity, stability and quantum efficiency, but they generally suffer from relatively long readout times. Unlike CCDs, which read out the image by continually transferring the charge between pixels towards a single amplifier and analog-to-digital converter (ADC), the newer CMOS sensors (developed since 1990s) employ individual amplifiers and ADCs for each column \cite{2020SPIE11454E..0GK}, which has dramatically reduced the readout times -- but at first at the expense of the other properties, so that CMOS chips originally found applications mostly within the realm of consumer electronics. Since 2009, cameras based on ``scientific CMOS'' sensors, with much improved performance have been available \cite{spie_scmos,spie_neo} and since then the variety of their manufacturers, as well as their sensitivity and size of the sensors has been increasing, while the costs have been decreasing, leading to a variety of applications in astronomy, in particular in applications requiring fast imaging, such as detection and study of rapid optical transients \cite{karpov_2010, karpov_2019}, space debris tracking \cite{karpov_2016} or observations of faint meteors \cite{karpov_meteors_2019}.  

\begin{figure}
\begin{center}
\includegraphics[width=37pc]{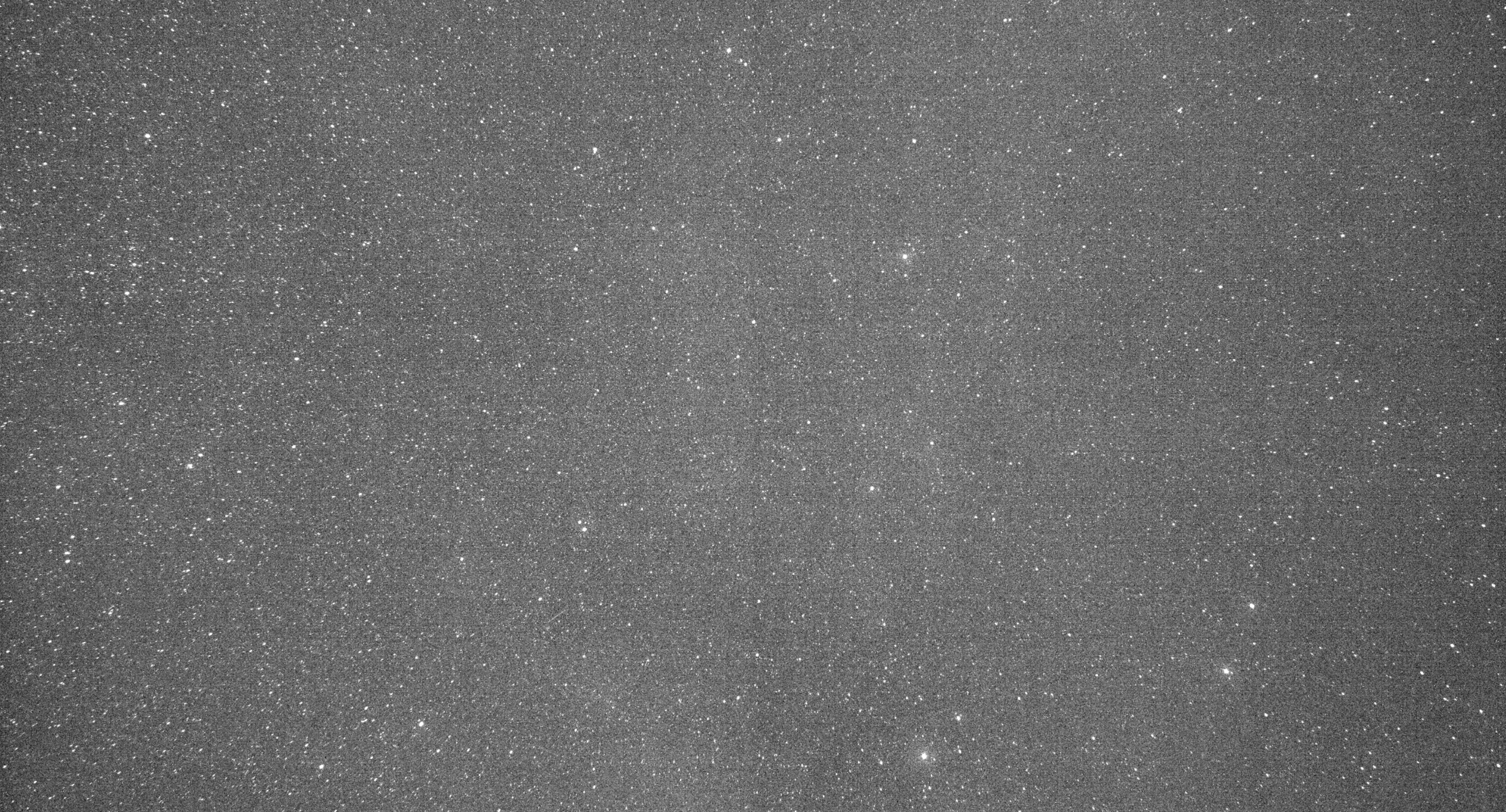}
\end{center}
\caption{\label{hvezdy}A test 3-second image from a 70/4 lens without any star tracking.}
\end{figure}

From the current spectrum of CMOS cameras, those with Sony IMX sensors are particularly common in astronomical applications \cite{2023PASP..135e5001A} (see references therein for examples) and thus well characterized. With sub-second readout times, these chips allow for a rapid mosaic coverage of a large part of the sky with a medium-focal-length lens. For the coverage of the field of view of the standard FD ($180\times28$ degrees), we consider as optimal the combination of IMX461 ($43.9\times32.9$ mm) and a 50--60 mm lens, which allows the coverage of the FD field of view in just 5 images. The high resolution of the chip leads to an image scale of 13 arcseconds per pixel, reducing the amount of blended stars in star-rich areas. As the sky on the equator moves by 15 arcseconds per second, 1-second exposures are possible without tracking at this image scale without significant blurring of star images. Longer exposures can be obtained by stacking the shorter exposures, since the readout time of the entire chip is only 0.66 seconds when implemented in the Moravian Instruments C5-100M camera and thus the time lost by readout is not excessive. Thus, the expensive and bulky equatorial mount can be completely omitted from the design.

\section{FramNG implementation}
For the optics, we preliminarily consider the Nikkor Z 58/0.95 lens, which is reported to be very sharp and also has a fully manual focusing mechanism, which is now uncommon, but highly desirable for astronomical applications. However the information on lens performance for star imaging are generally unreliable and we will select optics only after physical tests. Based on the extrapolation from our tests with different cameras and lenses (Figure~\ref{hvezdy}), we may be able to reach 11 magnitudes in zenith at f/0.95 already on 1-second exposures. However, we will optimize the combination of the f-stop (on which depends the PSF), the length of individual exposure (on which depends the motion blur) and possibly the amount of subsequent exposures to be stacked, if needed, based on the actual performance of the system,

We plan to mount the lens-camera combination on a simple horizontal turning platform and have it continuously scan the FD field of view -- even if three 1-second exposures are needed for each tile in the mosaic, the FD field of view will be covered fully in 25 seconds plus the time needed to turn the platform. An altitude movement mechanism will need to be added for the additional coverage of the field of view of the HEAT extension at  the Coihueco FD station. The lens-camera combination will weight about 5 kg and fit within a sphere of with a radius of 20 cm, so that the movement and enclosure (to protect it from inclement weather) can be much smaller and simpler than in the case of the traditional FRAMs. With the expected combined weight of less than 20 kg, the entire framNGs could be mounted on the roof of the FD buildings, removing the need for any additional infrastructure and further contributing to the lowering of the costs of the project.

A major challenge will be the processing of the data. Even if 3-second individual exposures are chosen, a single framNG will generate typically 1.5 TB of data per night. Such an amount of data is practical to neither store locally nor transmit away from the FD stations, meaning that all images will have to be immediately processed on site and most of them discarded afterwards. Based on the current experience with image processing, we expect that a standard PC with a 16-core CPU and 128 GB RAM will be able to handle the processing in real time, but reliable algorithms will have to be developed, if the raw data is not stored.

\section{Further applications}
Besides the primary goal of cloud detection, the large amount of imaging data gathered by framNGs may have a variety of further applications.

The large field of view of framNG in principle allows for precise aerosol determination using altitude fitting \cite{2021AJ....162....6E}. However if images are taken only in one altitude above the horizon, the extinction gradient due to changing airmass becomes degenerate with the aperture correction needed to account for the changes in the point spread function (PSF) across the image. This problem can be alleviated by using by taking images at different altitudes also at sites without the HEAT extension or by advancing the use of PSF correction in the analysis \cite{Negi_2022}. 

Currently, all-sky cameras are used by the operators of the Pierre Auger Observatory to evaluate sky brightness when the FD shutters are closed in order to take a decision to open the shutters. The data from framNGs could be used for the same purpose.

Besides cosmic ray showers, the FD can also detect various exotic atmospheric phenomena, in particular the elves \cite{2020E&SS....700582A}. The framNGs could record some related phenomena, since they are viewing the same field of view, however the practicability of this proposal is unclear and the real-time analysis would have to be capable of identifying such phenomena on the fly so that the relevant images are kept.

Finally, the large amount of measurements of stellar brightness could be used to search for variability, transient phenomena and other effects. Unlike the raw data, the measured brightness for each detected star can be feasibly recorded and the analysis can be done offline, provided that the behavior of the photometric procedure is well understood. Care must be taken to exclude the influence of stellar scintillation, which can be significant at such short time scales, as well as the effects due to the varying positions of the stars in the field of view. Additionally we consider periodically stacking series of consecutive raw images of the same direction, aligned to compensate for the rotation of the Earth, and storing the combined data in image form.

\section*{Acknowledgements}
This work was co-funded by the EU and supported by the Czech Ministry of Education, Youth and Sports through the projects  CZ.02.01.01/00/22\_008/0004632 and  CZ.02.01.01/00/22\_008/0004596.

{\footnotesize
\bibliography{ebr-atmohead24}{}

\providecommand{\newblock}{}
\begin{thebibliography}{10}
\expandafter\ifx\csname url\endcsname\relax
  \def\url#1{{\tt #1}}\fi
\expandafter\ifx\csname urlprefix\endcsname\relax\def\urlprefix{URL }\fi
\providecommand{\eprint}[2][]{\url{#2}}

\bibitem{PierreAuger:2015eyc}
Aab A {\em et~al.\/} (Pierre Auger Collaboration) 2015 {\em Nucl. Instrum. Meth. A\/} {\bf 798} 172--213 (\textit{Preprint} \eprint{1502.01323})

\bibitem{Chirinos}
{Chirinos, J} 2015 {\em EPJ Web of Conferences\/} {\bf 89} 03012

\bibitem{bianca}
Keilhauer B 2024 {\em EPJ Web of Conferences\/}  these proceedings

\bibitem{Blazek}
Blazek J 2017 {\em EPJ Web of Conferences\/} {\bf 144} 01009

\bibitem{2021JInst..16P6027A}
Aab A {\em et~al.\/} (Pierre Auger Collaboration) 2021 {\em Journal of Instrumentation\/} {\bf 16} P06027 (\textit{Preprint} \eprint{2101.11602})

\bibitem{2020SPIE11454E..0GK}
{Karpov} S, {Bajat} A, {Christov} A, {Prouza} M and {Beskin} G 2020 {\em X-Ray, Optical, and Infrared Detectors for Astronomy IX\/} ({\em SPIE Conference Series\/} vol 11454) p 114540G (\textit{Preprint} \eprint{2101.01517})

\bibitem{ABRAHAM2010227}
Abraham J {\em et~al.\/} (Pierre Auger Collaboration) 2010 {\em Nucl. Instrum. Meth. A\/} {\bf 620} 227--251

\bibitem{Mandat}
Mandat D {\em et~al.\/} 2015 {\em EPJ Web of Conferences\/} {\bf 89} 03007

\bibitem{Adam}
Adam J, Buss J~B, Brügge K, Nöthe M and Rhode W 2017 {\em EPJ Web of Conferences\/} {\bf 144} 01004

\bibitem{PierreAuger:2010swb}
Abreu P {\em et~al.\/} (Pierre Auger Collaboration) 2011 {\em Astropart. Phys.\/} {\bf 34} 368--381 (\textit{Preprint} \eprint{1010.6162})

\bibitem{2023A&A...674A...1G}
{Gaia Collaboration} {\em et~al.\/} 2023 {\em Astronomy and Astrophysics\/} {\bf 674} A1 (\textit{Preprint} \eprint{2208.00211})

\bibitem{2013AstBu..68..481Z}
{Zakharov} A~I, {Prokhorov} M~E, {Tuchin} M~S and {Zhukov} A~O 2013 {\em Astrophysical Bulletin\/} {\bf 68} 481--493

\bibitem{spie_scmos}
{Vu} P, {Fowler} B, {Liu} C, {Balicki} J, {Mims} S, {Do} H and {Laxson} D 2008 {Design of prototype scientific CMOS image sensors} {\em High Energy, Optical, and Infrared Detectors for Astronomy III\/} vol 7021 p 702103

\bibitem{spie_neo}
{Fowler} B, {Liu} C, {Mims} S, {Balicki} J, {Li} W, {Do} H, {Appelbaum} J and {Vu} P 2010 {\em Sensors, Cameras, and Systems for Industrial/Scientific Applications XI\/} ({\em SPIE Conference Series\/} vol 7536) p 753607

\bibitem{karpov_2010}
{Karpov} S, {Beskin} G, {Bondar} S, {Guarnieri} A, {Bartolini} C, {Greco} G and {Piccioni} A 2010 {\em Advances in Astronomy\/} {\bf 2010} 784141

\bibitem{karpov_2019}
{Karpov} S~o 2019 {\em Revista Mexicana de Astronomia y Astrofisica Conference Series\/} vol~51 p~30

\bibitem{karpov_2016}
{Karpov} S {\em et~al.\/} 2016 {\em Revista Mexicana de Astronomia y Astrofisica Conference Series\/} vol~48 pp 112--113

\bibitem{karpov_meteors_2019}
{Karpov} S {\em et~al.\/} 2019 {\em Revista Mexicana de Astronomia y Astrofisica Conference Series\/} vol~51 pp 127--130

\bibitem{2023PASP..135e5001A}
{Alarcon} M~R, {Licandro} J, {Serra-Ricart} M, {Joven} E, {Gaitan} V and {de Sousa} R 2023 {\em \pasp\/} {\bf 135} 055001 (\textit{Preprint} \eprint{2302.03700})

\bibitem{2021AJ....162....6E}
{Ebr} J, {Karpov} S, {Eli{\'a}{\v{s}}ek} J, {Bla{\v{z}}ek} J, {Cunniffe} R, {Ebrov{\'a}} I, {Jane{\v{c}}ek} P, {Jel{\'\i}nek} M, {Jury{\v{s}}ek} J, {Mand{\'a}t} D, {Ma{\v{s}}ek} M, {Pech} M, {Prouza} M and {Tr{\'a}vn{\'\i}{\v{c}}ek} P 2021 {\em \aj\/} {\bf 162} 6 (\textit{Preprint} \eprint{2101.03074})

\bibitem{Negi_2022}
Negi S, Ebr J, Karpov S and Eliášek J 2022 {\em Journal of Physics: Conference Series\/} {\bf 2398} 012019

\bibitem{2020E&SS....700582A}
{Aab} A {\em et~al.\/} (Pierre Auger Collaboration) 2020 {\em Earth and Space Science\/} {\bf 7} e00582

\end{thebibliography}
\bibliographystyle{iopart-num}}

\end{document}